\input colordvi
\documentclass[11pt]{amsart}
\usepackage{amsfonts,amssymb, graphicx,a4}
\usepackage{color}\definecolor{Red}{cmyk}{0,1,1,0}

\renewcommand{\qed}{\hfill\rule{3mm}{3mm}}
\newtheorem{teorema}{Theorem}
\newtheorem{proposition}[teorema]{Proposition}
\newtheorem{corollary}[teorema]{Corollary}
\newtheorem{Remark}[teorema]{Remark}
\newenvironment{remark}{\begin{Remark}\rm}{\end{Remark}}

\begin{document}


\voffset=-1.5truecm\hsize=16.5truecm    \vsize=24.truecm
\baselineskip=14pt plus0.1pt minus0.1pt \parindent=12pt
\lineskip=4pt\lineskiplimit=0.1pt      \parskip=0.1pt plus1pt

\def\ds{\displaystyle}\def\st{\scriptstyle}\def\sst{\scriptscriptstyle}


\let\a=\alpha \let\b=\beta
\let\d=\delta \let\e=\varepsilon
\let\f=\varphi \let\g=\gamma \let\h=\eta    \let\k=\kappa \let\l=\lambda
\let\m=\mu \let\n=\nu \let\o=\omega    \let\p=\pi \let\ph=\varphi
\let\r=\rho \let\s=\sigma \let\t=\tau \let\th=\vartheta
\let\y=\upsilon \let\x=\xi \let\z=\zeta
\let\D=\Delta \let\F=\Phi \let\G=\Gamma \let\L=\Lambda \let\Th=\Theta
\let\O=\Omega \let\P=\Pi \let\Ps=\Psi \let\Si=\Sigma \let\X=\Xi
\let\Y=\Upsilon

\global\newcount\numsec\global\newcount\numfor
\gdef\profonditastruttura{\dp\strutbox}
\def\senondefinito#1{\expandafter\ifx\csname#1\endcsname\relax}
\def\SIA #1,#2,#3 {\senondefinito{#1#2}
\expandafter\xdef\csname #1#2\endcsname{#3} \else
\write16{???? il simbolo #2 e' gia' stato definito !!!!} \fi}
\def\etichetta(#1){(\veroparagrafo.\veraformula)
\SIA e,#1,(\veroparagrafo.\veraformula)
 \global\advance\numfor by 1
 \write16{ EQ \equ(#1) ha simbolo #1 }}
\def\etichettaa(#1){(A\veroparagrafo.\veraformula)
 \SIA e,#1,(A\veroparagrafo.\veraformula)
 \global\advance\numfor by 1\write16{ EQ \equ(#1) ha simbolo #1 }}
\def\BOZZA{\def\alato(##1){
 {\vtop to \profonditastruttura{\baselineskip
 \profonditastruttura\vss
 \rlap{\kern-\hsize\kern-1.2truecm{$\scriptstyle##1$}}}}}}
\def\alato(#1){}
\def\veroparagrafo{\number\numsec}\def\veraformula{\number\numfor}
\def\Eq(#1){\eqno{\etichetta(#1)\alato(#1)}}
\def\eq(#1){\etichetta(#1)\alato(#1)}
\def\Eqa(#1){\eqno{\etichettaa(#1)\alato(#1)}}
\def\eqa(#1){\etichettaa(#1)\alato(#1)}
\def\equ(#1){\senondefinito{e#1}$\clubsuit$#1\else\csname e#1\endcsname\fi}
\let\EQ=\Eq


\def\\{\noindent}
\let\io=\infty

\def\VU{{\mathbb{V}}}
\def\EE{{\mathbb{E}}}
\def\GI{{\mathbb{G}}}
\def\TT{{\mathbb{T}}}
\def\C{\mathbb{C}}
\def\LL{{\cal L}}
\def\RR{{\cal R}}
\def\SS{{\cal S}}
\def\NN{{\cal N}}
\def\HH{{\cal H}}
\def\GG{{\cal G}}
\def\PP{{\cal P}}
\def\AA{{\cal A}}
\def\BB{{\cal B}}
\def\FF{{\cal F}}
\def\v{\vskip.1cm}
\def\vv{\vskip.2cm}
\def\gt{{\tilde\g}}
\def\E{{\mathcal E} }
\def\I{{\rm I}}

\def\cal{\mathcal}

\def\tende#1{\vtop{\ialign{##\crcr\rightarrowfill\crcr
              \noalign{\kern-1pt\nointerlineskip}
              \hskip3.pt${\scriptstyle #1}$\hskip3.pt\crcr}}}
\def\otto{{\kern-1.truept\leftarrow\kern-5.truept\to\kern-1.truept}}
\def\arm{{}}
\font\bigfnt=cmbx10 scaled\magstep1

\newcommand{\card}[1]{\left|#1\right|}
\newcommand{\und}[1]{\underline{#1}}
\def\1{\rlap{\mbox{\small\rm 1}}\kern.15em 1}
\def\ind#1{\1_{\{#1\}}}
\def\bydef{:=}
\def\defby{=:}
\def\buildd#1#2{\mathrel{\mathop{\kern 0pt#1}\limits_{#2}}}
\def\card#1{\left|#1\right|}
\def\proof{\noindent{\bf Proof. }}
\def\proofof#1{\noindent{\bf Proof of #1. }}
\def\trp{\mathbb{T}}
\def\trt{\mathcal{T}}

\def\bfz{\boldsymbol z}
\def\bfa{\boldsymbol a}
\def\bfalpha{\boldsymbol\alpha}
\def\bfmu{\boldsymbol \mu}
\def\bfmust{\bfT^\infty(\bfmu)}
\def\bfmupr{\boldsymbol {\widetilde\mu}}
\def\bfrho{\boldsymbol \rho}
\def\bfrhost{\boldsymbol \rho^*}
\def\bfrhopr{\boldsymbol {\widetilde\rho}}
\def\bfT{{\boldsymbol T}_{\!\!\bfrho}}
\def\bfR{\boldsymbol R}
\def\bfvarphi{\boldsymbol \varphi}
\def\bfvarphist{\boldsymbol \varphi^*}
\def\bfPi{\boldsymbol \Pi}
\def\bfzero{\boldsymbol 0}
\def\bfW{\boldsymbol W}
\def\formal{\stackrel{\rm F}{=}{}}
\def\eee{{\rm e}}
\def\nnn{\mathcal N}
\def\nst{\nnn^*}

\thispagestyle{empty}

\begin{center}
{\LARGE Cluster expansion for abstract polymer models.\\
New bounds from an old approach}
\vskip1.0cm
{\Large Roberto Fern\'andez$^{1}$ and Aldo Procacci$^{1,2}$}

\vskip.5cm

$^{1}$Labo. de Maths Raphael SALEM,
UMR 6085 CNRS-Univ. de Rouen, Avenue de l'Universit\'e, BP.12,
76801 Saint Etienne du Rouvray,  France\\
\vskip.1cm
 $^{2}$
Dep. Matem\'atica-ICEx, Universidade Federal de Minas Gerais, CP 702,
Belo Horizonte
MG 30.161-970, Brazil

\vskip.1cm
email: $^1${\tt Roberto.Fernandez@univ-rouen.fr};
~ $^2${\tt aldo@mat.ufmg.br}

\end{center}

\def\be{\begin{equation}}
\def\ee{\end{equation}}

\vskip1.0cm
\begin{quote}
{\small

  We revisit the classical approach to cluster expansions, based on
  tree graphs, and establish a new convergence
  condition that improves those by Koteck\'y-Preiss and
  Dobrushin, as we show in some examples.  The two ingredients
  of our approach are: (i) a careful consideration of the Penrose identity
  for truncated functions, and (ii) the use of iterated transformations
  to bound tree-graph expansions.

}
\end{quote}

\vskip1.0cm
\numsec=2\numfor=1
\\{\bf 1. Introduction}
\vv
\noindent
Cluster expansions, originally developed to express thermodynamic
potentials as power series in activities, are at the heart of
important perturbative arguments in statistical mechanics and other
branches of mathematical physics.  The classical approach to obtain
convergence conditions was based on combinatorial
considerations~\cite{mal80, sei82}, which were greatly simplified
through the use of tree-graph bounds~\cite{cam82, bry84}.  A
completely new inductive approach originated in the work of Koteck\'y
and Preiss \cite{kotpre86}, later refined by Dobrushin
\cite{dob96,dob96a} and many
others~\cite{narolizah99,bovzah00,mir00,sok01,uel04,scosok05}.  This later
approach is mathematically very appealing and, in its more elegant
version~\cite{dob96,scosok05}, it even disposes of any reference to power
series, becoming, in Dobrushin's words,  a ``no-cluster-expansion''
approach.  The combinatorial approach, however, kept its adepts who
reformulated it in a very clear and compact way~\cite{pfi91} and showed
how it can lead to bounds at least as good as those given by Koteck\'y
and Preiss~\cite{PS}.

In this paper, we revisit the classical combinatorial approach and
point out that it can be used, in a rather simple and natural way, to
produce improved bounds on the convergence region and the
sum of the expansion.  Our approach has two ingredients. First,
we exploit an \emph{identity},
due to Oliver Penrose~\cite{pen67}, relating the coefficients of the
expansion to a family of trees determined by compatibility constraints.
(As a matter of fact, we learnt this identity from the nice exposition in \cite[Section 3]{pfi91}.)
Successive approximations are obtained by
considering larger families of trees that neglect some of the
constraints. If only the very basic constraint is kept (links in the
tree must relate incompatible objects), the Kotecky-Preiss
condition emerges.  To the next order of precision (branches must end
in different objects) Dobrushin's condition is found.  By refining this
last constraint (branches' ends must be mutually compatible rather than
just different) we obtain a new convergence condition which
leads to improvements in several well-studied cases.   In particular,
for polymers on a graph ---for which compatibility means non-intersection---
our criterion yields the \emph{original} polymer condition due to
Gruber and Kunz~\cite[formula (42)]{grukun71}.
This somehow forgotten condition ---which is better than the ones usually applied---
was obtained in the very paper that introduced
the polymer formalism, through the use of Kirkwood-Salzburg equations.

Our second ingredient is a strategy to sum tree-graph expansions that
is complementary to the classical one.
The latter is based on an inductive ``defoliation'' of tree diagrams, which are summed
``from the leaves in'' with the help of the convergence condition.   Here, we show
instead that tree expansions are generated by successive applications of a
transformation defined by the convergence condition.
Besides leading to an improved convergence criterion,
this point of view presents, in our opinion, several advantageous features.
On the conceptual side, it shows a direct link between the convergence of
tree expansions and inequalities involving the functions found in
Koteck\'y-Preiss and Dobrushin (and our) conditions:  The inequalities
ensure that the iterative procedure lead to a finite expansion.
From a more practical point of view, it is easy to see that finite iterations of the transformations
yield progressively sharper bounds on the tree expansions.
Thus, our approach produces, for each convergence condition, an
associated sequence of upper bounds for the pinned free energy.
In particular the majorizing tree expansions are shown to be fix points of
the corresponding transformations.   All this information is absent
in previous treatments.

Finally, regarding future work, our approach leaves ample room
for extensions and improvements.  To emphasize this  fact, we
state a general result (Proposition \ref{prop:6}) showing how
bounds on truncated functions translate into convergence criteria
and associated results.  To establish our new criterion we used the
Penrose identity in the most natural and immediate way.  Improvements
should come from the incorporation of additional
tree conditions contained in the Penrose identity or, for specific models,
through a more accurate description of the compatibility constraints.
Also,  as emphasized in \cite{scosok05} and reviewed in Section 4.1,
there is a generalized Penrose identity which allows the use of trees
other than Penrose's to characterize truncated functions.  These alternative
choices may turn out to be of interest in particular settings.

Penrose identity, in its original or generalized form ---and thus our approach---
is valid only for hard-core
interactions (incompatibilities).  The extension of our treatment to polymer
systems subjected to  softer interactions is another direction for further
research.

\numsec=2\numfor=1
\vv\vv
\\{\bf 2. Set up and previous results}
\vv
\noindent
We adopt the following abstract polymer setting.  The
starting point is an unoriented graph $\GG=(\PP,\E)$ ---the
\emph{interaction graph}--- on a countable vertex set.  The vertices
$\g\in\PP$ are called \emph{polymers} for historical
reasons~\cite{grukun71}.  The name is misleading;
Dobrushin~\cite{dob96a} proposes to call them \emph{animals}, but the
traditional name holds on.  The edge set corresponds to an
\emph{incompatibility relation}: Two polymers $\g,\g'$ are
incompatible if $\{\g,\g'\}\in\E$, in which case we write
$\g\nsim\g'$.  Otherwise they are \emph{compatible} and we write
$\g\sim\g'$.
(Unfortunately, this notation ---well established within the
mathematical-physics community--- is the opposite to that
adopted in graph theory.)
The set of edges is arbitrary, except for the assumption
that it contains all pairs of the form $\{\g,\g\}$, that is,
\emph{every polymer is assumed to be incompatible with itself}.  In
particular vertices can be of infinite degree (each polymer can be
incompatible with infinitely many other polymers).  This happens, for
instance, for graphs associated to gases of low-temperature contours
or ``defects".

The physical information of each polymer model is given by the
incompatibility relation and a family of \emph{activities} $\bfz
=\{z_\g\}_{\g\in \PP}\in\C^{\PP}$.  For each \emph{finite} family
$\L\subset \PP$, these ingredients define probability weights on the
set of subsets of $\Lambda$:
$$ {\rm Prob}_\Lambda\bigl(\{\g_{1},\g_{2},\ldots,\g_{n}\}\bigr) \;=\;
\frac{1}{\Xi_{\Lambda}(\bfz)}\, z_{\g_{1}} z_{\g_{2}}\cdots z_{\g_{n}}
\prod_{j<k} \ind{\g_{j}\sim \g_{k}}
\Eq(r.1)
$$
for $n\ge 1$ and $ {\rm Prob}_\Lambda(\emptyset)=1/\Xi_{\L}$,
where
$$
\Xi_{\L}(\bfz)=1+\sum_{n\geq 1}{1\over n!}
\sum_{(\g_{1},\dots ,\g_{n})\in\L^n}
{z_{\g_1}}{z_{\g_2}}\dots{z_{\g_n}}
\prod_{j<k} \ind{\g_{j}\sim \g_{k}}\;.
\Eq(2)
$$

In physical terms, the measure \equ(r.1) corresponds to the
grand-canonical ensemble of a polymer gas with activities $\bfz$ and
hard-core interaction defined by the incompatibility relation.  The abstract
formalism makes it equivalent to a lattice gas on the graph $\GG$
with self- and nearest-neighbor hard-core repulsion. The
normalization constant \equ(2) is the grand-canonical partition function in the
``volume" $\L$.  Cluster expansions allow the control of the measures
\equ(r.1) uniformly in $\L$ and absolutely in the activities.  [Thus,
the control extends to the unphysical region of non-positive (complex)
activities, where the expressions on the right-hand side of \equ(r.1)
do not define probability measures.]
The basic cluster expansion is  the \emph{formal} power series (``F'') of the
logarithm of the partition function, which takes the form
(Mayer expansion, see e.g.~\cite{R})
$$
\log \Xi_{\L}(\bfz)\;\formal\; \sum_{n=1}^{\infty}{1\over n!}
\sum_{(\g_{1},\dots ,\g_{n})\subset\L^n}
\phi^{T}(\g_1 ,\dots , \g_n)\,z_{\g_1}\dots z_{\g_n}\Eq(6)
$$
with
$$
\phi^{T}(\g_{1},\dots ,\g_{n})\;=\;\left\{
\begin{array}{ll}
1&n=1\\[8pt]
\sum\limits_{G\subset \GG_{\{\g_{1},\dots ,\g_{n}\}}\atop G\ {\rm conn.\ spann.}}
 (-1)^{\card{E(G)}}&
n\ge 2\,,\;\GG_{\{\g_{1},\dots ,\g_{n}\}} \mbox{ connected}\\[25pt]
0 & n\ge 2\,,\; \GG_{\{\g_{1},\dots ,\g_{n}\}} \mbox{ not connected}
\end{array}\right.
\Eq(7)
$$
where $\GG_{\{\g_{1},\dots ,\g_{n}\}}$ is the graph of vertices
$\{1,\ldots,n\}$ and edges
$\bigl\{\{i.j\}: \g_i\nsim\g_j, 0\le i,j \le n\bigr\}$ and $G$
ranges over all its connected spanning subgraphs; here
$E(G)$ is the edge set of $G$.  The functions $\phi^{T}$ are the \emph{truncated functions}
of order $n$ (also called \emph{Ursell functions}).  The families
$\{\g_{1},\dots ,\g_{n}\}$ such that $\GG_{\{\g_{1},\dots ,\g_{n}\}}$ is connected
are the \emph{clusters}.

A telescoping argument shows that the properties of the measures
\equ(r.1) are determined by the one-polymer ratios (``pinned'' expansions)
$$
\Bigl[\log {\Xi_{\L}\over \Xi_{\L\setminus\{\g_0\}}}\Bigr](\bfz)
\;\formal\;
\sum_{n=1}^{\infty}{1\over n!}
\sum_{(\g_{1},\dots ,\g_{n})\subset\L^n\atop \exists i:\,\g_i=\g_0}
\phi^{T}(\g_1 ,\dots , \g_n)\,z_{\g_1}\dots z_{\g_n}
\Eq(6b)
$$
for each $\g_0\in\L$.  A more efficient alternative is to consider instead
the formal series

$$
\Bigl[{\partial\over\partial z_{\g_0}}\log\Xi_\L \Bigr](\bfz)\;\formal\; 1+
\sum_{n=1}^{\infty}{1\over n!}
\sum_{(\g_{1},\dots ,\g_{n})\subset\L^n}
\phi^{T}(\g_0,\g_1 ,\dots , \g_n)\,z_{\g_1}\dots z_{\g_n}\;.
\Eq(TP.0)
$$
This leads to the (infinite volume) formal  power series
$$
\Pi_{\g_0}(\bfrho)\;\bydef\; 1+
\sum_{n=1}^{\infty}{1\over n!}
\sum_{(\g_{1},\dots ,\g_{n})\in\PP^n}
\card{\phi^{T}(\g_0,\g_1 ,\dots , \g_n)}\,\r_{\g_1}\dots \r_{\g_n}\;,
\Eq(TP)
$$
for $\bfrho\in[0,\infty)^\PP$
---in which $\phi^T$ is replaced by $\card{\phi^T}$---
and its finite-volume versions $\Pi^\L_{\g_0}$
obtained by restricting the sum to polymers in $\L$.  The finiteness of the
positive-term series \equ(TP) for a certain $\bfrho$ implies the absolute
convergence of \equ(6), \equ(6b) and \equ(TP.0), uniformly in $\L$
for $|\bfz|\le\bfrho$.
This leads to the control of the measures \equ(r.1) and their $\L\to\PP$ limit~\cite{dob96a}.
[Throughout this paper, operations and relations involving boldface symbols
should be understood componentwisely, for instance $\bfrho\le\bfmu$ indicates
$\r_\g\le\mu_\g$, $\g\in\PP\,$; $-\bfz = \{-z_\g\}_{\g\in\PP}\,$;
$\bfrho\,\bfPi=\{\r_\g\Pi_\g\}_{\g\in\PP}\,$;
$\card{\bfz} = \{\card{z_\g}\}_{\g\in\PP}$, etc.]

The truncated functions satisfy the alternating-sign property
$$
\phi^{T}(\g_0,\g_1 ,\dots , \g_n) \;=\; (-1)^n\,
\card{\phi^{T}(\g_0,\g_1 ,\dots , \g_n)}\;.
\Eq(olv.1)
$$
(This is a well known result, that appears, for instance,
in~\cite[Theorem 4.5.3]{R} where it is
attributed to Groeneveld~\cite{gro62}.  Other proofs can be found in~\cite{mir00,scosok05}
and in Proposition \ref{prop:4} below).
Thus, \equ(TP.0) and the $\L$-restriction of \equ(TP) are related in the form
$$
\Pi^\L_{\g_0}(\bfrho) \;\formal\;
\Bigl[{\partial\over\partial z_{\g_0}}\log\Xi_\L \Bigr](-\bfrho)
\qquad (\bfrho\in[0,\infty)^\PP).
\Eq(olv.2)
$$
In the sequel we focus on the convergence of the series \equ(TP)
for positive activities. Its convergence allows the removal of the
label ``F'' in all precedent identities, and it implies the inequalities

$$
\biggl|\Bigl[{\partial\over\partial z_{\g_0}}\log\Xi_\L \Bigr](\bfz)\biggr| \;\le\;
\Bigl[{\partial\over\partial z_{\g_0}}\log\Xi_\L \Bigr](-\card{\bfz})\;=\;
\Pi^\L_{\g_0}(\card{\bfz})\le \Pi_{\g_0}(\card{\bfz})\;.
\Eq(cdg.2)
$$

and
$$
\biggl|\Big[|\log {\Xi_{\L}\over \Xi_{\L\setminus\{\g_0\}}}\Bigr](\bfz)\biggr|
\;\le\; -\Bigl[\log {\Xi_{\L}\over \Xi_{\L\setminus\{\g_0\}}}\Bigr](-\card{\bfz})
\; \le\; \card{z_{\g_0}}\,\Pi^\L_{\g_0}(\card{\bfz}) \le\; \card{z_{\g_0}}\,\Pi_{\g_0}(\card{\bfz})
\Eq(cdg.1)
$$
A  rather detailed study of different properties of these objects can be found
in~\cite{scosok05}.
\medskip

In the present general setting, two benchmark convergence conditions were
published in 1986~\cite{kotpre86} and 1996~\cite{dob96}.
For comparison purposes it is useful to write them
in the following form.  Suppose that for some $\bfrho\in[0,\infty)^\PP$ there
exists $\bfmu\in[0,\infty)^\PP$ such that
$$
\r_{\g_0}\; \exp\Bigl[\sum_{\g\nsim\g_0} \mu_{\g}\Bigr]
\;\le\; \mu_{\g_0} \quad \mbox{ (Koteck\'y-Preiss)}
\Eq(r.kp)
$$
or
$$
\r_{\g_0}\,\prod_{\g\nsim\g_0} \bigl(1 + \mu_\g\bigr)
\;\le\; \mu_{\g_0} \quad \mbox{ (Dobrushin)}
\Eq(r.dob)
$$
for each $\g_0\in\PP$.
[Please note that the sum and product over $\gamma$ here include $\gamma_0$, which
is always incompatible with itself.]
Then the power series \equ(TP) converges for such $\bfrho$
and, moreover,
$$
\r_{\g_0}\,\Pi_{\g_0}(\bfrho) \;\le\; \mu_{\g_0}
\Eq(cdg.3)
$$
for each $\g_0\in\PP$.

The reader may be more familiar with the following
forms of these conditions.  The change of variables
$\mu_\g=\r_\g\,{\eee}^{a_\g}$ shows that condition \equ(r.kp) is equivalent
to the existence of $\bfa\in[0,\infty)^\PP$ such that
$$
\sum_{\g:\g\nsim\g_0} \r_{\g}\,{\eee}^{a_\g}
\;\le\; a_{\g_0} \quad \mbox{ (Koteck\'y-Preiss)}
\Eq(r.kp.1)
$$
for each $\g_0\in\PP$, and \equ(cdg.3) becomes $\bfPi\le \eee^{\bfa}$.
The substitution $\mu_\g=\eee^{\alpha_\g}-1$, on the other hand,
makes \equ(r.dob)
equivalent to the existence of $\bfa\in[0,\infty)^\PP$ such that
$$
\r_{\g_0}\;\le\;  \Bigl(\eee^{\alpha_{\g_0}}-1\Bigr)\,
\exp\Bigl(-\sum_{\g:\g\nsim\g_0} \alpha_\g\Bigr)
\quad \mbox{ (Dobrushin)}
\Eq(r.dob.1)
$$
for each $\g_0\in\PP$.

The inequality
$$
\prod_{\g\nsim\g_0} \bigl(1 + \mu_\g\bigr)
\;\le\; \exp\Bigl[\sum_{\g\nsim\g_0} \mu_{\g}\Bigr]
\Eq(cdg.4.0)
$$
shows that Dobrushin condition is an improvement over Koteck\'y-Preiss'.
Nevertheless, the latter is particularly suited for some applications
(see, for instance, \cite{sok01}) and, furthermore, can be extended to
polymers  with soft self- and two-body interactions.
By contrast, the Dobrushin condition can be extended to systems
with soft two-body interaction~\cite{sok01} but requires hard-core
self-interaction.
Looking to
inequality \equ(cdg.4.0) we see that the difference between both criteria lies
in factors $\mu_\g$ at powers higher than two,  which are absent in the
left-hand-side.  A quick illustration of the consequences of this fact is provided
by polymers subjected only to self-exclusion
(each polymer is compatible with everybody else, except itself).  In this case
$\Xi_\L=\prod_{\g\in\L} (1+z_\g)$ and
$$
\Pi_{\g_0}(\bfrho)\;=\;\sum_{ n\ge 0} \rho_{\g_0}^n \;\formal\; \frac{1}{1-\rho_{\g_0}}\;.
\Eq(cdg.4)
$$
The Koteck\'y-Preiss condition requires the existence of $\bfmu>\bfzero$ such that
$\rho_{\g_0}\,{\rm e}^{\mu_{\g_0}} \,\le\, \mu_{\g_0}$ for each $\g_o\in\PP$, and this yields
a radius of convergence for $\Pi_{\g_0}$ equal to
$\sup_{\mu_{\g_0}} \mu_{\g_0}\,{\rm e}^{-\mu_{\g_0}} = {\rm e}^{-1}$.  Dobrushin condition,
on the other hand, provides the sharp estimate
$\sup_{\mu_{\g_0}} \,\mu_{\g_0}/(1+\mu_{\g_0}) =1$.

\numsec=3\numfor=1
\vv\vv
\\{\bf 3. Results}
\vv
\noindent
\\{\bf 3.1\ New convergence criteria}
\vv
\noindent
Our new criterion involves the grand-canonical partition functions
$\Xi_{\nst_{\g_0}}$, associated to the polymer families
$\nst_{\g_0}=\{\g\in\PP:\g\nsim\g_0\}$, $\g_0\in\PP$
($\GG$-neighborhood of $\g_0$, \emph{including} $\g_0$).  These functions,
defined in \equ(2), can also be written in the form
$$
\Xi_{\nst_{\g_0}}(\bfmu)=1+\sum_{n\geq 1} \,\frac{1}{n!}
\sum_{(\g_{1},\dots ,\g_{n})\in\PP^n\atop
\g_0\nsim\g_i\,,\, \g_i\sim\g_j\,,\, 1\le i ,j\le n}
{\mu_{\g_1}}{\mu_{\g_2}}\dots{\mu_{\g_n}}
\Eq(r.2)
$$
because compatible polymers are different.
Here is the practitioner's version of our criterion (a more detailed statement
is given in Theorem \ref{th:2} below).

\begin{teorema}\label{th:1}
Let   $\bfrho\in [0,\infty)^{\PP}$.  If there exists a $\bfmu\in
  [0,\infty)^{\PP}$ such that
$$
\r_{\g_0}\,\Xi_{\nst_{\g_0}}(\bfmu) \;\le\; \mu_{\g_0} \;,\quad
\forall \g_0\in\PP\;,
\Eq(r.3)
$$
then the series $\Pi_{\g_0}(\bfrho)$ [defined in \equ(TP)] converges for
such $\bfrho$ and satisfies $\r_{\g_0}\,\Pi_{\g_0}(\bfrho)\le\mu_{\g_0}$.
\end{teorema}
The inequality
$$
\Xi_{\nst_{\g_0}}(\bfmu) \;\le\; \prod_{\g\nsim\g_0} \bigl(1 + \mu_\g\bigr)
\Eq(r.3b)
$$
shows that condition \equ(r.3) is an improvement over Dobrushin's condition
---which in turns is an improvement over Koteck\'y-Preiss' condition.  The
improvement comes from the fact that only monomials involving \emph{mutually
compatible} polymers are allowed in the left-hand side.  Such improvement
comes, therefore, from two sources:
\begin{itemize}
\item[(I1)] In $\Xi_{\nst_{\g_0}}$ there are no monomials involving triangle diagrams in $\GG$,
namely pairs of neighbors of $\g_0$ that are themselves neighbors.
\item[(I2)]  In $\Xi_{{\cal N}^*_{\gamma_0}}$, the only monomial containing
$\mu_{\gamma_0}$ is $\mu_{\gamma_0}$ itself,
because $\g_0$ is incompatible with all other polymers in $\nst_{\g_0}$.
\end{itemize}
Improvement (I2) is present whichever the graph $\GG$, and makes inequality \equ(r.3b)
strict except for the non-interacting example discussed
circa \equ(cdg.4).   The terms corresponding to (I1) and (I2) can be neatly separated by writing
$$
\Xi_{\nst_{\g_0}}(\bfrho) \;=\; \rho_{\g_0} + \Xi_{\nnn_{\g_0}}(\bfrho)
\Eq(cdg.5)
$$
where $\nnn_{\g_0}=\nst_{\g_0}\setminus\{\g_0\}$ ($\Xi_\emptyset\bydef 1$).  Using a
bound similar to \equ(r.3b) but for $\Xi_{\nnn_{\g_0}}$ we obtain another
criterion ---halfway between ours and Dobrushin's--- which may
be useful in some settings.
\begin{corollary}\label{cor:0}
Let   $\bfrho\in [0,\infty)^{\PP}$.  If there exists a $\bfmu\in
  [0,\infty)^{\PP}$ such that
$$
\r_{\g_0}\,\Bigl[\mu_{\g_0}+\prod_{\g\nsim\g_0\atop \g\neq\g_0} \bigl(1 + \mu_\g\bigr)
\Bigr]\;\le\; \mu_{\g_0} \;,\quad
\mbox{(improved Dobrushin)}
\Eq(cdg.6)
$$
for all $\g_0\in\PP$, then the series $\Pi_{\g_0}(\r)$ converges for
such $\bfrho$ and satisfies $\r_{\g_0}\,\Pi_{\g_0}(\r)\le\mu_{\g_0}$.
\end{corollary}
Our condition \equ(r.3) coincides with \equ(cdg.6) for
triangle-free graphs $\GG$ (ex.\ trees, $\mathbb{Z}^d$), and it is
maximally better for complete (``triangle-full'') graphs.  This
and other examples will be analyzed below.
Summing up, available
convergence conditions are of the form
$$
\r_{\g_0} \,\varphi_{\g_0}(\bfmu) \;\le\; \mu_{\g_0}
\Eq(cdg.7)
$$
with
$$
\varphi_{\g_0}(\bfmu) \;=\; \left\{\begin{array}{ll}
\exp\Bigl[\sum_{\g\nsim\g_0} \mu_{\g}\Bigr] & \mbox{ (Koteck\'y-Preiss)}\\[10pt]
\prod_{\g\nsim\g_0} \bigl(1 + \mu_\g\bigr)  & \mbox{ (Dobrushin)}\\[10pt]
\mu_{\g_0}+\prod_{\g\nsim\g_0\atop \g\neq\g_0} \bigl(1 + \mu_\g\bigr)
& \mbox{(improved Dobrushin)}\\[10pt]
\Xi_{\nst_{\g_0}}(\bfmu)  & \mbox{ (ours)}
\end{array}\right.
\Eq(cdg.8)
$$
Each condition is strictly weaker than the preceding one except
for the facts that the improved Dobrushin condition coincides with
Dobrushin's if the polymers are non-interacting (only
self-excluding) and with our condition if $\GG$ does not include
any triangle diagram. The corresponding criteria yield information
on two issues: (i) regions of convergence, and  (ii) upper bounds
on each $\Pi_{\g_0}$.

Regarding the first issue, it is known that the region of absolute
convergence of cluster expansions has the properties of being a
``down-region''
---convergence for $\bfrho$ entails convergence for $\bfrhopr\le\bfrho$---
and log-convex.  The latter means that if the series converges for
$\bfrho$ and $\bfrhopr$ then it converges for
$\bfrho^\lambda\,\bfrhopr^{1-\lambda}$ for $0\le\lambda\le
1$~\cite{scosok05}.  It is reassuring to verify that these
properties also hold for the regions of validity of conditions
\equ(cdg.7)/\equ(cdg.8). Indeed, the ``down'' character is
obvious, and the log-convexity property is a consequence of the
following proposition.

\begin{proposition}\label{prop:0}
Suppose $0\le \lambda\le 1$ and let us denote
$$
\mathcal{R}_{\rm CD} \;=\;
\Bigl\{(\bfrho,\bfmu)\in [0,\infty)^\infty \times [0,\infty)^\infty \Bigm|
\mbox{condition  {\rm CD} is satisfied}\Bigr\} \;,
\Eq(cdg.9)
$$
where ``CD'' stand for each of the conditions in \equ(cdg.7)/\equ(cdg.8).   Then,
$$
(\bfrho,\bfmu)\,,\, (\bfrhopr,\bfmupr) \,\in \mathcal{R}_{\rm CD} \quad\Longrightarrow\quad
 \Bigl(\bfrho^\lambda\,\bfrhopr^{1-\lambda}\,,\,\bfmu^\lambda\,\bfmupr^{1-\lambda}\Bigr)
\,\in \mathcal{R}_{\rm CD}\;.
\Eq(cdg.10)
$$
\end{proposition}

\proof
Given the form \equ(cdg.7) of the conditions, we see that it is enough to prove that
$$
\varphi_{\g_0}(\bfmu)^\lambda \, \varphi_{\g_0}(\bfmupr)^{1-\lambda} \;\ge\;
\varphi_{\g_0}(\bfmu^\lambda\,\bfmupr^{1-\lambda})
\Eq(cdg.11)
$$
for each of the functions $\varphi_{\g_0}$ in \equ(cdg.8).  For the last three functions
this is a consequence of H\"older inequality in the form
$$
\Bigl(\sum_{i=1}^n a_i\Bigr)^\lambda \, \Bigl(\sum_{i=1}^n b_i\Bigr)^{1-\lambda}
\;\ge\; \sum_{i=1}^n a_i^\lambda\,b_i^{1-\lambda}
\Eq(cdg.12.0)
$$
($a_i, b_i\ge 0$, $i=1,\ldots,n$).  For the
Koteck\'y-Preiss function, \equ(cdg.11) is a consequence of the inequality
$ \lambda a + (1-\lambda) b \ge a^\lambda b^{1-\lambda}$, valid
for $a,b\ge 0$ (this is an elementary inequality,
see \cite[p.\ 112]{roy68}).  \qed
\medskip

Our results on the second issue (upper bound on $\bfPi$)  are
contained in the following strengthening of Theorem \ref{th:1}.  Its formulation
relies on the map iterates used in Section 4.2 to sum tree-graph expansions.
For each fixed $\bfrho\in[0,\infty)^\PP$ let us consider the map
$\bfT:[0,\infty)^\PP\longrightarrow[0,\infty]^\PP$ defined by
$$\bfT(\bfmu)\;\bydef\; \bfrho\,\bfvarphi(\bfmu)\;
\Eq(r.7.0)
$$
where $\bfvarphi$ is any of the functions defined in \equ(cdg.8).  Denote
$\bfT^n=\bfT(\bfT^{n-1})$ the successive compositions of $\bfT$ with itself.

\begin{teorema}\label{th:2}
Let $\bfrho\in[0,\infty)^\PP$ be fixed and let $\bfT$ be a map of the form
\equ(r.7.0)/\equ(cdg.8).  Assume there
exists $\bfmu\in[0,\infty)^\PP$ satisfying \equ(cdg.7), that is,
$$\bfT(\bfmu) \;\le\; \bfmu \;.
\Eq(r.10)
$$
Then:
\begin{itemize}
\item[(i)] There exists $\bfrhost\in[0,\infty)^\PP$ such that
$\bfT^n(\bfrho)\nearrow \bfrhost$ and $T(\bfrhost)=\bfrhost$.
\item[(ii)] For each $n\in\mathbb{N}$,
$$  \bfrho\, \bfPi\;\le\; \bfrhost \;\le\;\bfT^{n+1}(\bfmu)\;\le\;\bfT^n(\bfmu)
\;\le\;\bfmu\;.
\Eq(r.12.x)
$$
\end{itemize}
\end{teorema}
The deepest statement in this theorem is the first inequality in \equ(r.12.x).  The
rest of the theorem follows from the fact that
for all choices \equ(cdg.8) of $\bfvarphi$ the map $\bfT$ is monotonicity-preserving
and satisfies $\bfrho\;\le\; \bfT(\bfrho)\;\le\;\bfT(\bfmu)\;\le\;\bfmu$.

\vv
\vv
\noindent
\\{\bf 3.2 \ Comparison with previous criteria}
\vv
\noindent
To test our criterion we compare the estimates of the regions of convergence
provided by the criteria \equ(cdg.7)--\equ(cdg.8)
for two families of benchmark examples.
\medskip

\paragraph{\bf Polymer graphs with bounded maximum degree}
These are examples where $\GG$ has maximum degree $\Delta<\infty$.  We
shall suppose that all polymers have equal activity $\r_\g\equiv\r$
for all $\g\in\GG$, and therefore we search for equally constant
functions $\mu_\g\equiv\mu$.  The preceding criteria take the form
$\r\le \mu/\varphi(\mu)$ for appropriate functions $\varphi$, and the
maximization of the right-hand side with respect to $\mu$ yields the
best lower bounds of the radius of convergence of \equ(TP) [and hence
of \equ(6)].

In Table \ref{tab:1a} we summarize both convergence criteria and
best estimates on the convergence radii obtained with Koteck\'y-Preiss,
Dobrushin and improved Dobrushin conditions.
The only feature of the graph $\GG$
relevant for these criteria is the maximal degree $\Delta$ of the vertices.
Therefore they provide the sharpest results for graphs which lack of any
other feature and whose vertices have all degree $\Delta$.  These are
the regular trees with branching rate $\Delta-1$.  This fact ---trees
supply a worst-case condition that can be used whenever we ignore,
or decide to ignore, any topological information on the graph--- has
been emphasized in~\cite{sok01} (see, also, Remark~\ref{rem:2}).
For regular trees, the weak
Dobrushin condition coincides with ours, and there is a further, optimal
condition, due to Scott and Sokal~\cite{scosok05}, which we have included
in the last line of the table.  This condition is derived through a sequence of
volume-dependent Dobrushin conditions.  It would be interesting to see
whether a similar strategy could be developed within our approach.

\begin{table}[h]
\renewcommand{\arraystretch}{1.7}
\begin{center}
\begin{tabular}{|c||c|c|c|}\hline
Condition & Criterion & $R_\Delta$ & $R_6$ \\
\hline\hline
Koteck\'y-Preiss &
\rule[-5mm]{0mm}{11mm}$\displaystyle \r\le \mu\,e^{-(\Delta+1)\mu}$ &
$\displaystyle {1\over (\D+1)\,e}$ & 0.0525\\
\hline
Dobrushin & $\displaystyle \r\le \frac{\mu}{(1+\mu)^{\Delta+1}}$ &
\rule[-8.5mm]{0mm}{19mm}$\displaystyle \left\{\begin{array}{ll}
\displaystyle {\D^\D\over (\D+1)^{\D+1}} & \Delta \ge 1\\
1\;(*) &\Delta=0\end{array}\right.$ & 0.0566\\
\hline
\begin{tabular}{l}
improved Dobrushin\\
=\protect\equ(r.3) for $\scriptstyle(\Delta-1)$-reg.~tree
\end{tabular} & $\displaystyle \r\le \frac{\mu}{\mu + (1+\mu)^{\Delta}}$ &
\rule[-9.5mm]{0mm}{21mm}$\displaystyle \left\{\begin{array}{ll}
\displaystyle\left[1+{\D^\D\over (\D-1)^{\D-1}}\right]^{-1}& \Delta \ge 2\\[8pt]
\displaystyle(\Delta+1)^{-1}\;(*) &\Delta=0,1\end{array}\right.$ & 0.0628\\
\hline
Scott-Sokal & \cite[Theorem 5.6]{scosok05} &
\rule[-4mm]{0mm}{11mm}$\displaystyle\frac{(\D-1)^{(\D-1)}}{ \D^\D} \;(*)$ & 0.067\\
\hline
\end{tabular}
\end{center}
\caption{Convergence criteria and lower bounds ($R_\Delta$)
on the radius of convergence
when $\GG$ is a graph with maximal degree $\Delta$.  A star indicates that the
value is exact for the $(\Delta-1)$-regular tree}
\label{tab:1a}
\end{table}

In Table \ref{tab:2a} we show the improved results obtained from the application
of our criteria to some popular examples.  The values of $R$ in the first
two lines are to be compared with the values for $R_6$ in Table \ref{tab:1a}, and
that of the complete graph with the values of $R_\Delta$.
The source of these improvements is, of course, the sensitivity of our new criterium
to triangle diagrams.  In particular, our criterion gives the exact
value of the radius of convergence for the complete graph, for which
 $\Pi=[1-(\Delta+1)\mu]^{-1}$.

\begin{table}[h]
\renewcommand{\arraystretch}{1.7}
\begin{center}
\begin{tabular}{|c||c|c|}\hline
Model & Criterion & $R$\\
\hline\hline
\rule[-5.5mm]{0mm}{13mm}Domino in $\mathbb{Z}^2$
& $\displaystyle\r\le \frac{\mu}{1+7\mu+9\mu^2}$
& 0.0769\\
\hline
\rule[-5.5mm]{0mm}{13mm} Triangular lattice
& $\displaystyle \r\le \frac{\mu}{1+7\mu+8\mu^2+2\mu^3}$
& $\displaystyle 4R^3+8R^2=1\;,\; R\approx 0,078$\\
\hline
\rule[-5.5mm]{0mm}{13mm} $\scriptstyle (\Delta+1)$-complete graph
& $\displaystyle \r\le \frac{\mu}{1+(\Delta+1)\mu}$
& $\displaystyle (\Delta+1)^{-1}\;(*)$\\
\hline
\end{tabular}
\end{center}
\caption{Convergence criteria and lower bounds ($R$)
on the radius of convergence obtained with condition
\protect\equ(r.3) for some graphs $\GG$ of finite degree.
A star indicates an exact value}
\label{tab:2a}
\end{table}

\medskip

\paragraph{\bf Polymers on a graph}
This is the general example of cluster expansions for
graphs with vertices of infinite degree.  Applications include contour ensembles of
low-temperature phases, geometrical objects of high-temperature
expansions, random sets of the Fortuin-Kasteleyn representation of the
Potts model, \dots The general setup for these models is a polymer
family formed by the finite parts of a given set $\VU$  with incompatibility
defined by overlapping.  (Usually, $\VU$ is formed by the vertices
of a graph with respect to which polymers form connected sets.)

For these systems it is useful and traditional to pass to exponential
weight functions $a(\g)$ defined by $\m_\g=\r_\g\,{\eee}^{a(\g)}$.
Condition \equ(r.3) becomes
$$
1+\sum_{n\geq 1}
\sum_{\{\g_{1},\dots ,\g_{n}\}\subset\PP\atop
\g_0\cap\g_i\neq\emptyset\,,\,
\g_i\cap\g_j=\emptyset\,,\, 1\le i ,j\le n}
\prod_{i=1}^n\r_{\g_i}\, \eee^{a(\g_i)}
\;\le\; \eee^{a(\g_0)}\;
\Eq(r.6)
$$
From the constraint in the sum we only keep the fact that each of the
polymers $\g_1,\ldots,\g_n$ must intersect \emph{different} points in $\g_0$
(otherwise they would overlap).  This implies: (i) $n\le \card{\g_0}$, and (ii)
there are $n$ different points in $\g_0$ touched by $\g_1\cup\cdots\cup\g_n$.
These points can be chosen in ${\card{\g_0}\choose n}$ ways.  Hence,
the left-hand side of \equ(r.6) is less or equal than
$$
1+ \sum_{n=1}^{\card{\g_0}}
{\card{\g_0}\choose n}\Bigg[\sup_{x\in \g_0}\,\sum_{\g\in  \PP\atop \g\ni x}
{\r_{\g}}\,\eee^{a(\g)}\Bigg]^n\;=\;
\Bigg[1+\sup_{x\in \g_0}\,\sum_{\g\in  \PP\atop \g\ni x}
{\r_{\g}}\,\eee^{a(\g)}\Bigg]^{\card{\g_0}}\;,
$$
which leads us to the following sufficient condition for \equ(r.6):
$$
\sup_{x\in \g_0}\,\sum_{\g\in  \PP\atop \g\ni x}
{\r_{\g}}\,\eee^{a(\g)}\;\le\; \eee^{a(\g_0)/\card{\g_0}}-1
\Eq(usht1)
$$
This condition entails the finiteness of $\bfPi$:
$$
\Pi_{\g_0}(\bfrho)\;\le\;{\eee}^{a(\g_0)}\;.
\Eq(cdg.12)
$$

In practice, the function $a(\g)$ is chosen of the form
$a(\g)=a\card\g$, with $a$ a positive constant.
This choice, which in many cases is the expected optimal asymptotic behavior
of $a(\g)$ for large polymers,
simplifies the procedure reducing it to the determination of the single constant $a$.
Our emphasis in a general dependence is not just mathematical finesse.
As dominants contributions come from the smallest polymers,
a dependence of $a(\g)$ dealing more accurately with them would improve precision.
Also, the criteria are
usually presented in the slightly weaker form obtained by replacing
the supremum over $x\in\g_0$ by a supremum over $x\in\VU$.  In this form,
a condition like \equ(usht1) is, in fact, present in the seminal paper
by Gruber and Kunz~\cite{grukun71} [formula (42) with
normalization $\phi(x)=1$ and parametrization $\xi_0=e^a$.]
Table \ref{tab:3} lists the different conditions with the preceding
usual choices.

\begin{table}[h]
\renewcommand{\arraystretch}{1.7}
\begin{center}
\begin{tabular}{|c|c|c|}\hline
Koteck\'y-Preiss & Dobrushin  & Gruber-Kunz \\
\hline\hline
\rule[-6.5mm]{0mm}{11mm}
$\displaystyle\sup_{x}\,\sum_{\g\in  \PP: \g\ni x}
{\r_{\g}}\,{\eee}^{a\card{\g}}\;\le\;  a$
&$\displaystyle\sup_{x}\,\prod_{\g\in  \PP: \g\ni x}
\Bigl[1+ {\r_{\g}}\,{\eee}^{a\card{\g}}\Bigr] \;\le\;  {\eee}^a$
& $\displaystyle\sup_{x}\,\sum_{\g\in  \PP: \g\ni x}
{\r_{\g}}\,{{\eee}^{a\card{\g}}\;\le\; {\eee}^a}-1$\\
\hline
\end{tabular}
\end{center}
\caption{Convergence conditions for general polymer models.  Our
condition \equ(usht1) with $a(\g)=a\card\g$ coincides with
that by Gruber and Kunz.}
\label{tab:3}
\end{table}

\numsec=4\numfor=1
\vv\vv
\\{\bf 4. Proofs}
\vv
\\
\noindent
The argument has two distinct parts.  First, we use
the Penrose tree identity for the truncated functions to turn \equ(TP)
into a sum over trees ---a \emph{tree-graph expansion}.  In the second
part, we control this expansion through a natural iterative
procedure defined by the functions \equ(cdg.8).

\vv\vv
\\{\bf 4.1 Partitionability and the Penrose identity}
\vv
\\
\noindent
Formula \equ(7) involves a huge number of cancellations.
Penrose~\cite{pen67} realized that they can be optimally handled
through what is now known as the property of  \emph{partitionability}
of the family of connected spanning subgraphs.
While his original argument involved a particular partition scheme,
it works equally well for any other choice, as emphasized in~\cite{scosok05}.
For the sake of completeness, and due to its potential use for extensions
and alternative versions of our criterion, we
start by reproducing this simple but deep argument.
Our exposition is based on~\cite[Section 2.2]{scosok05}.

Let us consider a finite graph $\mathbb{G}=(\mathbb{U}, \mathbb{E})$ and denote
$\mathcal{C}_{\mathbb G}$ the set of all connected spanning subgraphs of $\mathbb{G}$
and $\mathcal{T}_{\mathbb G}$ the family of trees belonging to $\mathcal{C}_{\mathbb G}$.
Further, we consider $\mathcal{C}_{\mathbb G}$ partial ordered by bond inclusion:
$$G\le \widetilde G \quad \Longleftrightarrow\quad E(G) \subset E(\widetilde G)\;.
\Eq(cdg.13)
$$
If $G\le \widetilde G$, let us denote $[G,\widetilde G]$ the set of
$\widehat G\in\mathcal{C}_{\mathbb G}$
such that $G\le \widehat G\le \widetilde G$.  Let us call a
\emph{partition scheme} for the family $\mathcal{C}_{\mathbb G}$
to any map $R:\mathcal{T}_{\mathbb G} \to \mathcal{C}_{\mathbb G}$
such that
\begin{itemize}
\item[(i)] $E\bigl(R(T)\bigr)\supset E(T)$, and
\item[(ii)] $\mathcal{C}_{\mathbb G}$ is the disjoint union of the sets
$[T,R(T)]$, $T\in\mathcal{T}_{\mathbb G}$.
\end{itemize}

A number of such partition schemes are by now available (see references
in~\cite[Section 2.2]{scosok05}).  The one proposed by Penrose
is constructed in the following way:  Let us fix an enumeration
$v_0, v_1,\ldots, v_n$ for the vertices of $\mathbb{G}$, and
for each $\tau\in\mathcal{T}_{\mathbb{G}}$
(thought as a tree rooted in $v_0$),
let $d(i)$ be the tree distance of the vertex $v_i$ to $v_0$.
Penrose scheme associates to $\tau$ the graph $R_{\rm Pen}(\tau)$
formed by adding (only once) to $\tau$
all edges $\{v_i,v_j\}\in \mathbb{E}\setminus E(\tau)$ such that either:
\begin{itemize}
\item[(p1)]  $d(i)=d(j)$ (edges between vertices of the same generation), or
\item[(p2)]  $d(i)=d(j)-1$ and $i<j$ (edges connecting to predecessors with smaller index).
\end{itemize}

For a partition scheme $R$, let us denote
$$\mathcal{T}_{R} \;\bydef\;\Bigl\{ \tau\in \mathcal{T}_{\mathbb{G}}
\Bigm| R(\tau)=\tau \Bigr\}
\Eq(cdg.14)
$$
(set of $R$\emph{-trees}).  In particular, $ \mathcal{T}_{R_{\rm Pen}}$
is the set of \emph{Penrose trees}.
The following is the generalized version of Penrose identity.
\begin{proposition}\label{prop:4}
$$
\sum_{G\in\mathcal{C}_{\mathbb G}} (-1)^{\card{E(G)}} \;=\;
(-1)^{\card{\mathbb{V}}-1} \bigl| \mathcal{T}_{R} \bigr|\;.
\Eq(cdg.15)
$$
for any partition scheme $R$.
\end{proposition}

\proof
For any numbers $x_e$, $e\in \mathbb{E}$, we have
$$
\begin{array}{rcl}
\displaystyle\sum_{G\in\mathcal{C}_{\mathbb G}} \prod_{e\in E(G)} x_e
&=& \displaystyle\sum_{T\in\mathcal{T}_{\mathbb G}} \prod_{e\in E(T)} x_e
\sum_{\mathcal{F}\subset E(R(T))\setminus E(T)}
\,\prod_{e\in \mathcal{F}} x_e\\[20pt]
\displaystyle&=& \displaystyle\sum_{T\in\mathcal{T}_{\mathbb G}} \prod_{e\in E(T)} x_e
\prod_{e\in E(R(T))\setminus E(T)} (1+x_e)\;.
\end{array}
\Eq(cdg.16)
$$
The first equality is due to property (ii) of partition schemes.
If $x_e=-1$, the last factor kills the contributions of all trees with
$E(R(T))\setminus E(T)\neq\emptyset$.  Furthermore, for any tree
$\bigl|E(T)\bigr|=\card{\mathbb{V}}-1$.\qed
\medskip

We see that the hard-core condition is crucial for the identity.  For
polymer models with soft repulsion, only
$\card{1+x_e}\le 1$ is guaranteed, and this leads to the inequality
$$
\Bigl|\sum_{G\in\mathcal{C}_{\mathbb G}} \prod_{e \in E(G)} x_e\Bigr|
\;\le\;\sum_{T\in\mathcal{T}_{\mathbb G}} \prod_{e\in E(T)} \card{x_e}
\;\le\; \card{\mathcal{T}_{\mathbb G}}\;.
\Eq(cdg.17)
$$
This much weaker inequality is the one used in traditional treatments of the
tree expansion~\cite{cam82, bry84}.

The previous proposition applied to the Penrose scheme implies
$$
\card{\phi^{T}(\g_0,\g_{1},\dots ,\g_{n})}\;=\;
\sum_{T\in\mathcal{T}_n^0}
\ind{T\in \mathcal{T}_{\rm Pen}(\g_0,\g_1 ,\dots , \g_n)}\;,
\Eq(TP.P)
$$
where $\mathcal{T}_n^0$ is the set of (labeled) trees with vertices $\{0,1,\ldots,n\}$
rooted in 0, and
$\mathcal{T}_{\rm Pen}(\g_0,\g_1 ,\dots , \g_n)$ denotes  the Penrose
trees on the graph $\GG_{\{\g_0,\g_{1},\dots ,\g_{n}\}}$ (with the canonical
enumeration of vertices).  Similar formulas are valid replacing ``Pen" by
any partition scheme $R$.

\begin{remark}\label{rem:2}
  As the number of Penrose trees grows with the disappearance of
  triangle diagrams, the value of $\bfPi$ (resp.\ the region of
  convergence of the cluster expansion) for a given graph $\GG$ is
  bounded above by (resp.\ contains) that of a tree where each vertex
  has a degree larger or equal than that at $\GG$.  Furthermore, the
  latter is bounded above (resp.\ contains) that of a homogeneous tree
  with branching rates equal to the maximal rate.
\end{remark}

\vv\vv
\\{\bf 4.2 Trees and convergence}
\vv
\\
\noindent
Replacing \equ(TP.P) into \equ(TP) we obtain a sum in terms of trees.
Traditionally, such expansions have been inductively summed
\emph{alla} Cammarota~\cite{cam82}, namely ``from the leaves in".
Conditions of the type \equ(cdg.7) guarantee the reproducibility of the inductive
hypothesis.
Here we present a complementary approach, based in
generating the expansion through repeated
application of a nonlinear map $\bfT$.  Conditions \equ(cdg.7) prevent
the successive partial sums to diverge.

The end product of this section
is the following proposition.  Each $\tau\in\mathcal{T}_n^0$ is uniquely
defined by the branching factor $s_i$ of each vertex $i$ and the labels
$i_1,\ldots, i_{s_i}$ of its descendants.

\begin{proposition}\label{prop:6}
Let $\GG=(\PP,\E)$ be a polymer system and assume
there exist functions $c_n:\PP^{n+1}\to [0,\infty)$, for
$n\in\mathbb{N}$, invariant under permutations of
the last $n$ arguments such that
$$
\card{\phi^{T}(\g_0,\g_1 ,\dots , \g_n)}\;\le\;
\sum_{\tau \in\trt^0_n}
\,\prod_{i=0}^n
c_{s_i}(\g_i,\g_{i_1},\ldots,\g_{i_{s_i}})\;.
\Eq(r.16)
$$
Consider the function  $\bfvarphi:[0,\infty)^\PP\to [0,\infty]^\PP$
defined by
$$
\varphi _{\g_0} (\bfmu)\;=\; 1+\sum_{n\geq 1} \frac{1}{n!}\,
\sum_{(\g_{1},\dots ,\g_{n})\in\PP^n} c_n(\g_0,\g_1,\ldots,\g_n)\,
{\mu_{\g_1}}\dots{\mu_{\g_n}}
\Eq(r.15.2)
$$
for each $\g_0\in\PP$.
Assume that, for a given $\bfrho\in[0,\infty)^\PP$ there exists
$\bfmu\in[0,\infty)^\PP$ such that
$$
\r_{\g_0}\,\varphi_{\g_0}(\bfmu)\;\le\;\ \mu_{\g_0}
\Eq(r.17)
$$
for each $\g_0\in\PP$.  Then,
\begin{itemize}
\item[(a)] The cluster expansion \equ(6)
for the system $\GG$ converges absolutely and uniformly in $\Lambda$
and in the activities $\bfz$ with $\card{\bfz} \le\bfrho$.
\item[(b)] Furthermore, if $\bfT=\bfrho\,\bfvarphi$ is the map defined as in \equ(r.7.0) but with
$\bfvarphi$ given by \equ(r.15.2), then
\begin{itemize}
\item[(i)] There exist $\bfrhost,\bfmust \in [0,\infty)^\PP$ such that
$$
\bfT^n(\bfrho)\buildd{\nearrow}{n\to\infty} \bfrhost \quad,\quad
\bfT^n(\bfmu)\buildd{\searrow}{n\to\infty} \bfmust \;.
\Eq(olv.10)
$$
\item[(ii)] $T(\bfrhost)=\bfrhost$.
\item[(iii)] For each $n\in\mathbb{N}$,
$$\bfrho\, \bfPi\;\le\; \bfrhost \;\le\; \bfmust \;\le\;\bfT^{n+1}(\bfmu)\;\le\;\bfT^n(\bfmu)
\;\le\;\bfmu\;.
\Eq(r.12)
$$
\end{itemize}
\end{itemize}
\end{proposition}

The proof requires only elementary manipulations which,
however, require some previous
considerations to introduce the necessary notation.

It is useful to visualize the maps \equ(r.7.0) in the diagrammatic form
\setlength{\unitlength}{1cm}
$$
\begin{picture}(12,2.5)
\thicklines
\put(0,1.5){$\Bigl(\bfT(\bfmu)\Bigr)_{\g_0}$}
\put(2,1.5){=}
\put(2.7,1.5){$\circ$}
\put(2.7,1.3){$\scriptstyle\g_0$}
\put(3.15,1.5){$+$}
\put(3.7,1.5){$\circ$}
\put(3.7,1.3){$\scriptstyle\g_0$}
\put(3.85,1.58){\line(1,0){1}}
\put(4.8,1.5){$\bullet{\scriptstyle 1}$}
\put(5.3,1.5){$+$}
\put(5.8,1.5){$\circ$}
\put(5.8,1.3){$\scriptstyle\g_0$}
\put(5.97,1.58){\line(2,1){0.8}}
\put(6.67,1.89){$\bullet{\scriptstyle 1}$}
\put(5.97,1.58){\line(2,-1){0.8}}
\put(6.67,1.08){$\bullet{\scriptstyle 2}$}
\put(7.12,1.5){$+$}
\put(7.6,1.5){$\cdots$}
\put(8.25,1.5){$+$}
\put(8.8,1.5){$\circ$}
\put(8.8,1.3){$\scriptstyle\g_0$}
\put(8.97,1.58){\line(1,1){0.7}}
\put(9.64,2.23){$\bullet{\scriptstyle 1}$}
\put(8.97,1.58){\line(2,1){0.8}}
\put(9.64,1.86){$\bullet{\scriptstyle 2}$}
\put(9.64,1.3){\vdots}
\put(8.97,1.58){\line(1,-1){0.7}}
\put(9.64,0.74){$\bullet{\scriptstyle n}$}
\put(10.09,1.5){$+$}
\put(10.57,1.5){$\cdots$}
\end{picture}
$$
The sum is over all single-generation rooted trees.  In each tree,
open circles represents a factor $\r$, bullets a factor $\mu$ and
vertices other than the root must be summed over all possible polymers
$\g$.  At each vertex with $n$ descendants, a ``vertex function'' $c_n/n!$
acts, having as arguments the ordered $n+1$-tuple formed by the polymer
at the vertex, the polymer at the top offspring, the polymer at the
next offspring from the top,\ldots, in that order.
With this representation, the iteration $T^2(\bfmu)$ corresponds to
replacing each of the bullets by each one of the diagrams of the
expansion for $T$.  This leads to rooted trees of up to two
generations, with open circles at first-generation vertices and bullets
at second-generation ones.  In particular, all single-generation trees
have only open circles.  Notice that the two drawings of Figure
\ref{fig:1} appear in two different terms of the expansion, and hence
should be counted as \emph{different} diagrams.  More generally, the
$k$-th iteration of $T$ involves all possible rooted tree diagrams,
counting as different those obtained by permutations of non-identical
branches.  We shall call these diagrams \emph{planar rooted trees}.
In each term of the expansion, vertices of generation $k$ are occupied
by bullets and all the others by open circles.

Formally, the definition of planar rooted trees is determined by a
labeling choice which we fix as follows.  There is a special vertex,
labeled $0$ (the root), placed, say, at the leftmost position of the
drawing.  From it there emerge $s_{0}$ branches ending at the
first-generation vertices.  The value $s_{0}=0$ describes the trivial
tree with the root as its only vertex.  Otherwise these vertices are
drawn along a vertical line at the right of the root and labeled
$(0,1),\dots (0,s_{0})$ with the second subscript increasing from the
top to the bottom of the line.  The construction continues rightwards:
Each of the vertices $(0,i)$, gives rise to a family of
second-generation vertices $(0,i,1),\dots (0,i,s_{(0,i)})$ and so on.
The vertex $v$ is of generation $\ell$ if its label has the form
$v=(0,i_1,\ldots,i_\ell)$ with $1\le i_j\le s_{(0,i_1,\ldots,i_{j-1})}$,
$1\le j\le \ell$ ($i_0\equiv0$).  The sequence of such branching factors
$s_{(0,i_1,\ldots,i_\ell)}\in\mathbb{N}\cup\{0\} $ define the
planar rooted tree.  Let us denote $\trp^{0,k}$ the
set of trees with maximal generation number $k$; $\trp^{0,0}$ being
the trivial tree.  Figure \ref{fig:1} shows two different trees of
$\trp^{0,2}$.  We enumerate the vertices following the generation number
and the ``top to bottom'' order in case of equal generation.  [This amounts
to declaring $(0,i_1,\ldots,i_\ell)<(0,i'_1,\ldots,i'_{\ell'})$ if
$\ell< \ell'$ and using lexicographic order if $\ell= \ell'$].

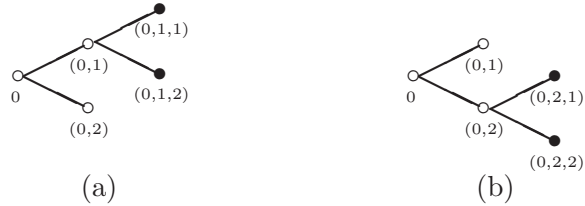
\begin{figure}[h]
\begin{center}
\begin{picture}(3,3)
\thicklines
\put(0,1.5){$\circ$}
\put(0,1.25){$\scriptscriptstyle 0$}
\put(0.17,1.59){\line(2,1){0.8}}
\put(0.93,1.93){$\circ$}
\put(0.77,1.68){$\scriptscriptstyle (0,1)$}
\put(0.17,1.59){\line(2,-1){0.8}}
\put(0.93,1.07){$\circ$}
\put(0.77,0.82){$\scriptscriptstyle (0,2)$}
\put(1.12,2.05){\line(2,1){0.8}}
\put(1.88,2.39){$\bullet$}
\put(1.63,2.15){$\scriptscriptstyle (0,1,1)$}
\put(1.12,2.05){\line(2,-1){0.8}}
\put(1.88,1.53){$\bullet$}
\put(1.63,1.29){$\scriptscriptstyle (0,1,2)$}
\put(0.93,0){(a)}
\end{picture}
\hspace{2cm}
\begin{picture}(3,3)
\thicklines
\put(0,1.5){$\circ$}
\put(0,1.25){$\scriptscriptstyle 0$}
\put(0.17,1.59){\line(2,1){0.8}}
\put(0.93,1.93){$\circ$}
\put(0.77,1.68){$\scriptscriptstyle (0,1)$}
\put(0.17,1.59){\line(2,-1){0.8}}
\put(0.93,1.07){$\circ$}
\put(0.77,0.82){$\scriptscriptstyle (0,2)$}
\put(1.12,1.16){\line(2,1){0.8}}
\put(1.88,1.5){$\bullet$}
\put(1.63,1.25){$\scriptscriptstyle (0,2,1)$}
\put(1.12,1.16){\line(2,-1){0.8}}
\put(1.88,0.64){$\bullet$}
\put(1.63,0.39){$\scriptscriptstyle (0,2,2)$}
\put(0.93,0){(b)}
\end{picture}
\end{center}
\caption{Planar rooted trees defined by
(a) $s_0= s_{(0,1)}=2$ and $s_{(0,2)}=s_{(0,1,1)}=s_{(0,1,2)}=0$;
(b) $s_0= s_{(0,2)}=2$ and $s_{(0,1)}=s_{(0,2,1)}=s_{(0,2,2)}=0$}
\label{fig:1}
\end{figure}

A straightforward inductive argument shows that
$$
\Bigl(\bfT^k(\bfmu)\Bigr)_{\g_0} \;=\; \r_{\g_0} \Bigl[\sum_{\ell=0}^{k-1}
R^{(\ell)}_{\g_0}(\bfrho) + R^{(k)}_{\g_0}(\bfrho,\bfmu)\Bigr]
\Eq(r.8.1)
$$
with
$$
R^{(\ell)}_{\g_0}(\bfrho) \;=\; \,\sum_{t \in\trp^{0,\ell}}\,
\,\sum_{(\g_{v_1},\ldots ,\g_{v_{\card{V_t }}})\in\PP^{\card{V_t }}}\,
\prod_{i=0}^{\card{V_t}} \frac{1}{s_{v_i}!}\,
c_{s_{v_i}}(\g_{v_i},\g_{({v_i},1)},\ldots,\g_{({v_i},s_{v_i})})\,
\r_{\g_{({v_i},1)}}\dots\r_{\g_{({v_i},s_{v_i})}}
\Eq(r.9)
$$
and $R^{(k)}_{\g_0}(\bfrho,\bfmu)$ has a similar expression but
with the activities of the vertex of the $k$-th generation
weighted by $\bfmu$.  In this expression $V_t$ denotes the set of
non-root vertices of $t$ and we agree that $c_0(\g_v)\equiv 1$
and $\prod_\emptyset \equiv 1$.
We are interested in the $k\to\infty$ limit
of \equ(r.8.1).  Let us denote $\TT^0=\cup_\ell\TT^{0,\ell}$.
These considerations make almost immediate the proof of the
following lemma which, together with a simple combinatorial argument,
proves Proposition \ref{prop:6}.

\begin{proposition}\label{prop:1}
For some fixed $\bfrho\in[0,\infty)^\PP$ let
$\bfT$ be a map of the form \equ(r.7.0)/\equ(r.15.2) and assume there
exists $\bfmu\in[0,\infty)^\PP$ such that  $\bfT(\bfmu) \;\le\; \bfmu$.
Then $\bfT^n(\bfrho)\nearrow \bfrhost \in [0,\infty)^\PP$ as $n\to\infty$, with
$$
\r^*_{\g_0} \;\bydef\; \r_{\g_0}
\,\sum_{t \in\trp^0}
\,\sum_{(\g_{v_1},\ldots ,\g_{v_{\card{V_t }}})\in\PP^{\card{V_t }}}\,
\prod_{i=0}^{\card{V_t}} \frac{1}{s_{v_i}!}\,
c_{s_{v_i}}(\g_{v_i},\g_{({v_i},1)},\ldots,\g_{({v_i},s_{v_i})})\,
\r_{\g_{({v_i},1)}}\dots\r_{\g_{({v_i},s_{v_i})}}
\Eq(r.11)
$$
for each $\g_0\in\PP$.  Furthermore,
\begin{itemize}
\item[(i)] $T(\bfrhost)=\bfrhost$.
\item[(ii)] There exists $\bfmust \in [0,\infty)^\PP$ such that
$\bfT^n(\bfmu)\searrow \bfmust$ as $n\to\infty$.
\item[(iii)] For all $\ell,n\in\mathbb{N}$,
$$ \bfmu\;\ge\; \bfT^n(\bfmu)
\;\ge\; \bfT^{n+1}(\bfmu) \;\ge\;
\bfmust \;\ge\; \bfrhost
\;\ge\; \bfT^{\ell+1}(\bfrho) \;\ge\; \bfT^{\ell}(\bfrho) \;\ge\; \bfrho\;.
\Eq(r.12.n)
$$
\end{itemize}
\end{proposition}

\proof  The map $\bfT$ is obviously monotinicity preserving in the
coordinatewise partial order of $[0,\infty]^\PP$
and
$$
\bfmu\;\ge\; \bfT(\bfmu)\;\ge\;\bfT(\bfrho)\;\ge\;\bfrho\;.
\Eq(olv.11)
$$
[The first inequality is by hypothesis, the second one by monotonicity and
the  third one is immediate from the definition of $\bfT$.]
Therefore, by induction,
$$
\bfmu\;\ge\; \bfT^{n}(\bfmu)\;\ge\; \bfT^{n+1}(\bfmu)
\;\ge\;\bfT^{n+\ell+1}(\bfmu)\;\ge\;
\bfT^{n+\ell+1}(\bfrho)\;\ge\;\bfT^{\ell+1}(\bfrho)
\;\ge\;\bfT^\ell(\bfrho)\;\ge\;\bfrho
\Eq(r.13)
$$
for all $\ell,n\in\mathbb{N}$.  This shows that, for each
$\g\in\GG$, the series $\bigl(\bfT^\ell(\bfrho)\bigr)_\ell$ is increasing and bounded
above while $\bigl(\bfT^n(\bfmu)\bigr)_n$ is decreasing and bounded below.
Thus, the limits
$\bfrhost\bydef\sup_\ell \bfT^\ell(\bfrho)$ and
$\bfmust\bydef\inf_\ell \bfT^\ell(\bfmu)$ exist and are
finite and, by letting alternatingly $\ell\to\infty$
and $n\to\infty$ in \equ(r.12), we obtain the inequalities
\equ(r.12.n).
The fact that $\bfT^\infty(\bfrho)=\bfrho^*$ is immediate from expression
\equ(r.8.1).  Finally,
$$ \bfrhost\;=\;\lim_{n\to\infty}\bfT\bigl(\bfT^n(\bfrho)\bigr) \;=\;
\bfT\Bigl(\lim_{n\to\infty}\bfT^n(\bfrho)\Bigr)\;=\; \bfT(\bfrhost)
\Eq(fin.1)
$$
where the middle identity is by monotone convergence.  $\qed$
\bigskip

We notice that
$T^{\infty}_{\g_0}(\bfmu)=\r^*_{\g_0} + \lim_k R^{(k)}_{\g_0}(\bfmu)$.
The last limit is in fact an infimum because
$\bfR^{(k)}(\bfrho,\bfmu)\le \bfR^{(k-1)}\bigl(\bfrho,\bfT(\bfmu)\bigr)\le
\bfR^{(k-1)}(\bfrho,\bfmu)$.
\bigskip

\proofof{Proposition \ref{prop:6}}  The sum in \equ(r.11) can be
written in the form
$$
\r^*_{\g_0} \;=\; \r_{\g_0} \,\sum_{t \in\trp^0} W_{\g_0}(t)\;.
\Eq(fin.2)
$$
The symmetry of the vertex functions $c_n(\g_0,\g_1,\ldots, g_n)$ implies
that the weights $\bfW(t)$ that are
invariant under permutations of the branches of the planar tree $t$.
That is, they depend only on the underlying labeled tree $\tau$ obtained by
neglecting the order of the vertices.  Formally, if $\trt^0_n$ is the
set of rooted trees
on $\{0,1,\ldots,n\}$ (=labelled trees of $n+1$
vertices), there is a map $\trt^0_n\ni \tau\mapsto t_\tau\in\trp^0_n$
where $t_\tau$ is the planar tree obtained by drawing branches
starting on the root according to the order given by the labels of the
first offspring, and continuing in this way for branches within
branches.  This map is many-to-one, in fact, the cardinality of the
preimage of a tree $t$
(=number of ways of labelling the $\card{V_t}$ non-root vertices of a
planar rooted tree with $\card{V_t}$ distinct labels
consistently with the rule ``from high to low")
is
$$
\b_t\;=\; \frac{\card{V_t}!}{\prod_{i=0}^{\card{V_t}} s_{v_i}!}\Eq(rel1)
$$
(see e.g. theorem 145B in \cite{but87}).
Thus, we can replace the sum in \equ(fin.2) by a sum over trees $\tau$
on the set $\trt^0=\cup_n \trt^0_n$ of rooted trees:
$$
\r^*_{\g_0} \;=\; \r_{\g_0} \,\sum_{\tau\in\trt^0}
\frac{W_{\g_0}(t_\tau)}{\b_{t_\tau}}\;.
\Eq(fin.3)
$$
If we expand $\bfW$ and permute the sum over trees with the sum over
polymer sequences (allowed operation for a series of positive terms), we obtain
$$
\r^*_{\g_0} \;=\; \r_{\g_0}\,\sum_{n\ge 0} \;\frac{1}{n!}
\,\sum_{(\g_{1},\ldots ,\g_n)\in\PP^n}\,
\Bigl[\sum_{\tau \in\trt^0_n} \,
\prod_{i=0}^n  c_{s_i}(\g_i,\g_{i_1},\ldots,\g_{i_{s_i}})\Bigr]
\; \r_{\g_1} \cdots\r_{\g_n}\;.
\Eq(r.15.1b)
$$
Comparing this expression with \equ(TP), we see immediately
that hypothesis \equ(r.16) implies that $\bfrho\,\bfPi\le \bfrhost$.
The remaining statements are a consequence of Proposition \ref{prop:1}.
\qed

\vv\vv
\\{\bf 4.2 Proof of Theorem \ref{th:2}}
\vv
\\
\noindent
We just have to show that the different convergence conditions can be written in
the form \equ(r.15.2) for vertex functions $c_s$ satisfying \equ(r.16).  Theorem
\ref{th:2} then follows from Proposition \ref{prop:6}.

We use Penrose identity \equ(TP.P) to obtain a bound of the form \equ(r.16).
For this, we keep only the vertex constraints of a Penrose tree $\tau$:  The descendants
of a given vertex may not be linked by an edge in the initial graph $\GG$.  Otherwise
[by condition (p1) in Sectionn 4.1],
the graph $R_{\rm Pen}(\tau)$ would include such an edge and would, therefore,
differ from $\tau$.  That is, we consider the  larger family of trees such that
 $$ \mbox{If $\{i,i_1\}$ and $\{i,i_2\}$ are edges of $\tau$, then $\g_{i_1}\sim\g_{i_2}$}
 \Eq(r.20b)
 $$
In this way we obtain bounds of the form \equ(r.16) with
$$
c_n(\g_0,\g_1,\ldots,\g_n) \;=\;
\prod_{i=1}^n \ind{\g_0\nsim\g_i} \prod_{j=1}^n \ind{\g_i\sim\g_j}\;,
\Eq(r.21)
$$
and Proposition \ref{prop:6} applies with
$$
\varphi _{\g_0} (\bfmu)\;=\; 1+\sum_{n\geq 1} \,\frac{1}{n!}
\sum_{(\g_{1},\dots ,\g_{n})\in\PP^n\atop
\g_0\nsim\g_i\,,\, \g_i\sim\g_j\,,\, 1\le i ,j\le n}
{\mu_{\g_1}}\dots{\mu_{\g_n}}\;,=\; \Xi_{\PP_{\g_0}}(\bfmu)\;.
\Eq(r.22)
$$
This proves the criterion of Theorem \ref{th:1}.

If we replace in \equ(r.20b) the condition $\g_i\nsim\g_j$ by the
weaker requirement $\g_i\neq\g_j$ we obtain
$$
c^{\rm Dob}_n(\g_0,\g_1,\ldots,\g_n) \;=\;
\prod_{i=1}^n \ind{\g_0\nsim\g_i} \prod_{j=1}^n \ind{\g_i\neq\g_j}\;,
\Eq(r.24)
$$
and
$$
\varphi^{\rm Dob} _{\g_0} (\bfmu)\;=\; 1+\sum_{n\geq 1}
\frac{1}{n!}\,\sum_{(\g_{1},\dots ,\g_{n})\in\PP^n\atop
\g_0\nsim\g_i\,,\, \g_i\neq\g_j\,,\, 1\le i ,j\le n}
{\mu_{\g_1}}\dots{\mu_{\g_n}}\;=\;
\prod_{\g\nsim\g_0} (1+\mu_\g)\;,
\Eq(r.25)
$$
which corresponds to Dobrushin condition.   The improved Dobrushin
condition is obtained by strengthening \equ(r.24) through the
further requirement that $\g_i\neq\g_0$ for $i=1,\ldots,n$ and $n\ge 2$.

Finally, if requirement \equ(r.20b) is ignored altogether,
$$
c^{\rm KP}_n(\g_0,\g_1,\ldots,\g_n) \;=\;
\prod_{i=1}^n \ind{\g_0\nsim\g_i} \;,
\Eq(r.26)
$$
and
$$
\varphi^{\rm KP} _{\g_0} (\bfmu)\;=\; 1+\sum_{n\geq 1} \,\frac{1}{n!}
\sum_{(\g_{1},\dots ,\g_{n})\in\PP^n\atop
\g_0\nsim\g_i\,,\,  1\le i \le n}
{\mu_{\g_1}}\dots{\mu_{\g_n}}\;=\;
\exp\Bigl[\sum_{\g\nsim\g_0} \mu_{\g}\Bigr]
\Eq(r.27)
$$
yields the criterion of Koteck\'y and Preiss. \qed

\section*{Acknowledgements}
We are indebted to the two referees
for detailed and constructive criticism that prompted us to a
substantial rewriting of the introductory sections and a
clearer layout of our proof.  We are thankful to
Benedetto Scoppola, Roman Koteck\'y and Alan Sokal for long and
fruitful discussions which led us to a number of notions that have
enriched our presentation (in particular, partitionability
of complexes of edges, log-convexity of convergence regions and the
fixed-point character of $\bfrhost$).
The authors thank Warwick University and the Universit\`a di Roma
``Tor Vergata'' (AP) for hospitality during these
discussions.   It is a pleasure to thank Aernout van
Enter for useful comments and for a critical reading of
the manuscript, and Charles-Edouard Pfister and Daniel Ueltschi
for encouragement.  The work of AP was supported by a visitor grant of
CAPES (Coordena\c{c}\~ao de Aperfei\c{c}oamento de Pessoal de
N\'{\i}vel Superior, Brasil).  He also thanks the Mathematics
Laboratory Raphael Salem of the University of Rouen for the invitation
that started the project and for hospitality during its realization.
The work of RF was partially supported by the project GIP-ANR
NT05-3-43374 (Agence Nationale de la Recherche, France).

\end{document}